\documentclass[]{spie}  %>>> use for US letter paper
\usepackage{amsmath,amsfonts,amssymb}
\usepackage{mathrsfs,empheq}
\usepackage{mathtools}
\usepackage{graphicx}
\usepackage[colorlinks=true, allcolors=blue]{hyperref}
\usepackage{cite}

\title{Early Detection of Retinopathy of Prematurity stage using Deep Learning approach}

\author[a]{Supriti Mulay}
\author[b]{Keerthi Ram}
\author[a,b]{Mohanasankar Sivaprakasam}
\author[c]{Anand Vinekar}
\affil[a]{Healthcare Technology Innovation Centre, Chennai, India}
\affil[b]{Dept. of Electrical Engg., IIT Madras, Chennai, India}
\affil[c]{Narayana Nethralaya Postgraduate Institute of Ophthalmology, Bangalore, India}

\authorinfo{Further author information: (Send correspondence to Supriti Mulay.)\\Supriti Mulay.: E-mail: supriti@htic.iitm.ac.in}

\begin{document} 
\maketitle

\begin{abstract}
Retinopathy of Prematurity (ROP) is a fibrovascular proliferative disorder, which affects the developing peripheral retinal vasculature of premature infants. Early detection of ROP is possible in stage 1 and stage 2 characterized by demarcation line and ridge with width which separates vascularised retina and the peripheral retina. To detect demarcation line/ ridge from neonatal retinal images is a complex task because of low contrast images. In this paper we focus on detection of ridge, the important landmark in ROP diagnosis, using Convolutional Neural Network(CNN). Our contribution is to use a CNN-based model Mask R-CNN for demarcation line/ridge detection allowing clinicians to detect ROP stage 2 better. The proposed system applies a pre-processing step of image enhancement to overcome poor image quality. In this study we use labelled neonatal images and we explore the use of CNN to localize ridge in these images. We used a dataset of 220 images of 45 babies from the KIDROP project. The system was trained on 175 retinal images with ground truth segmentation of ridge region. The system was tested on 45 images and reached detection accuracy of 0.88, showing that deep learning detection with pre-processing by image normalization allows robust detection of ROP in early stages.  
\end{abstract}

% Include a list of keywords after the abstract 

\section{INTRODUCTION}
\label{sec:intro}  % \label{} allows reference to this section

Retinopathy of prematurity (ROP) is a disorder of the developing retina of low-birth-weight preterm infants that potentially leads to blindness in a small but significant percentage of those infants. ROP can be mild with no visual defects, or it may become aggressive with new blood vessel formation (neovascularization) and progress to retinal detachment and blindness. There are five stages of ROP, from mild (Stage 1) to severe (Stage 5) in which the retina detaches in the eye. Babies with Stage 1 and 2 ROP are called prethreshold, and those with Stages 3 through 5 are called threshold. In determining the ROP stages, ridge plays a critical role as the site of vascular shunting. Retinal photography may improve the objective documentation of disease findings, increase the accuracy and standardization of diagnosis and eventually improve the accessibility, cost, efficiency, and safety of ROP care through telemedicine [\citen{Chiang}]. Early diagnosis of damage is important in the treatment of ROP. Therefore detection of stage1 and stage 2 ROP is crucial.
		
In order to provide diagnostic assistance to ophthalmologist for ROP stage detection, automatic detection of demarcation line/ridge pattern plays a significant role. Demarcation line is definite structure that separates the avascular retina anteriorly from the vascularized retina posteriorly. Demarcation line develops into ridge with width and height. Figure. 1 shows some of the examples of demarcation line/ridge along with enhanced images. For the early detection and effective treatment of ROP, demarcation line detection is challenging due to insufficient understanding of ROP symptomatology, lack of gold-standard ground-truth data and poor quality fundus imaging.

\begin{figure} [ht]
   \begin{center}
   \begin{tabular}{c} %% tabular useful for creating an array of images 
   \includegraphics[height=5cm]{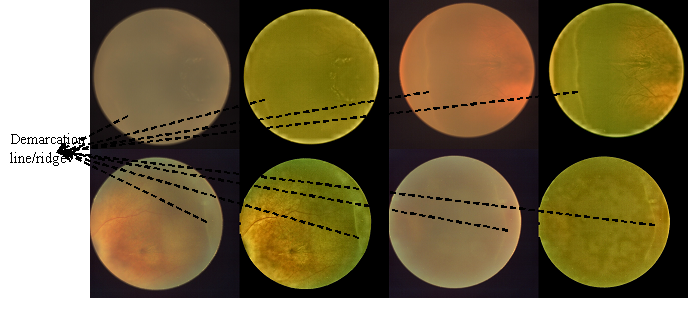}
   \end{tabular}
   \end{center}
   \caption[example] 
%>>>> use \label inside caption to get Fig. number with \ref{}
   { \label{fig:example1} 
demarcation line/ridge examples.}
   \end{figure}  

We believe we are the first to propose ridge detection using CNN for early ROP staging. [\citen{SinthanyC}] applied image processing watershed technique to detect ridge in neonatal images. But along with ridge other regions which are not the ROP also got detected. Indigenous software [\citen{JoshiM}] for automatic detection of stages of ROP uses multilevel vessel enhancement to enhance tubular structures in the images to identify the ridge/ demarcation line. In contrast to this we use Mask R-CNN to detect ridge and able to localize the demarcation line/ridge with state of the art performance. This method localizes ridge or demarcation line, but distinction between stage 1 and stage 2 ROP is difficult, as there is a very thin line between them with this method. 

\section{METHOD}

Our ROP ridge detection consists of a deep learning pipeline: preprocessing, feature extractor (CNN) and fine tuning layers. Figure. 2 shows the end-to-end pipeline of this process. 

\begin{figure} [ht]
   \begin{center}
   \begin{tabular}{c} %% tabular useful for creating an array of images 
   \includegraphics[height=6cm]{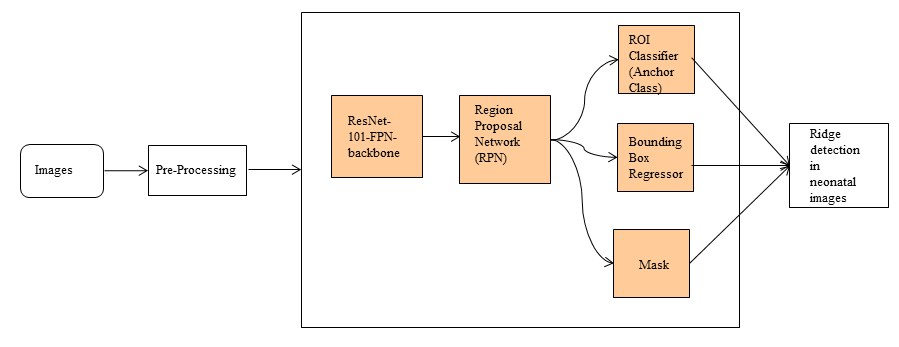}
   \end{tabular}
   \end{center}
   \caption[example] 
%>>>> use \label inside caption to get Fig. number with \ref{}
   { \label{fig:example2} 
End to end pipeline for detecting ridge.}
   \end{figure}  

Typically, there are three steps in an object detection framework. 
\begin{enumerate}
\item First, a model or algorithm is used to generate regions of interest or region proposals. These region proposals are a large set of bounding boxes spanning the full image (i.e. an object localisation component (ResNet-101-FPN and RPN)). 
\item In the second step, visual features are extracted for each of the bounding boxes, they are evaluated and it is determined whether, and which objects are present in the proposals based on visual features (i.e. an object classification component ROI Classifier).
\item In the final post-processing step, overlapping boxes are combined into a single bounding box (i.e. non maximum suppression (NMS)).
\end{enumerate}

\subsection{Data annotations}
\label{sec:title}
For this study we used images from the KIDROP project[\citen{kidrop}]. The experimental dataset consisted of 220 neonatal images of 45 babies. The ground truth for ridge in each image was determined manually by an independent annotator. The annotator depicted bounding boxes surrounding each ridge. These manual labels were used for developing the models and evaluating their performance. The initial size of some images was 2040x2040 and for some 1020x1020, was reduced to 1024x800 to avoid the overhead introduced by upsampling the predicted masks to the original image resolution. 

We marked the boundary of demarcation line for ground truth image creation. The ground truth images are verified by clinicians.

\subsection{Pre-processing/normalization}
Uneven illumination, blurring, incorrect focus, and low contrast reduce the quality of neonatal retinal images, resulting in a loss of sensitivity and specificity for diagnostic purposes, and may even impair the ophthalmologist's ability to interpret significant eye features or distinguish different retinal diseases[\citen{SevikU}]. Poor quality retinal images make it difficult for subsequent accurate segmentation and computer-aided diagnosis of retinal diseases,which are used to automate the detection process and to assist ophthalmologists[\citen{ZhouM}]. As the quality of fundus images is poor, image enhancement technique is applied to get the "primal sketch" of retinal image. The color images have been enhanced by modified histogram equalization (MHE) algorithm [\citen{Kareem}]. On the lightness component in YIQ color space (Y component), the adaptive histogram equalization (CLAHE) method is applied. In order to preserve the mean brightness of the input image, the parameters of a sigmoid function are chosen to minimize the absolute mean brightness metric. Image transformation using sigmoid function is shown in the following equations.

\begin{equation}
    { f(i,j)\rightarrow  g(i,j) = 255 * (\frac{\psi (f)-\psi (fmin)}{\psi (fmax)-\psi (fmin)})}
\end{equation}
where $\psi (f)=[1+exp(c*(0.05-f))]^{-1}$ is the Sigmoid function, 
      $f$  - input image ,  $g$  - output image and   $c$ - sigmoid parameter, set to 2.5.
      
 After CLAHE, blur effect occurs in the image. To remove the additive noise and invert the blurring simultaneously, Wiener filtering is performed. The demarcation line that is crucial in determining the ROP stage is prominently enhanced using this method. Enhanced ridges are shown in Figure. 3.

\begin{figure} [ht]
   \begin{center}
   \begin{tabular}{c} %% tabular useful for creating an array of images 
   \includegraphics[height=3cm]{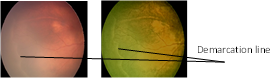}
   \end{tabular}
   \end{center}
   \caption[example] 
%>>>> use \label inside caption to get Fig. number with \ref{}
   { \label{fig:example3} 
Enhanced Retinal Image with YIQ Adapt HIST Method.}
   \end{figure} 

\subsection{Demarcation line/ridge Localization}
Localization of demarcation line/ridge is done using Mask R-CNN. Mask R-CNN  (regional convolutional neural network)[\citen{KaimingGD}] is a two stage framework: the first stage scans the image and generates proposals (areas likely to contain an object). This stage is a light-weight neural network called Region proposal network (RPN) which scans all feature pyramid network (FPN) top-bottom pathway and proposes regions which may contain objects. The second stage classifies the proposals and generates bounding boxes and masks. This stage is another neural network which takes regions proposed by the first stage and assigns them to several specific areas of a feature map level, scans these areas, and generates objects classes, bounding boxes and masks. Faster R-CNN is extended to Mask R-CNN by adding a branch to predict segmentation masks for each region of interest generated in Faster R-CNN. Mask R-CNN is a state-of-the-art architecture for ridge detection. 

\begin{figure} [ht]
   \begin{center}
   \begin{tabular}{c} %% tabular useful for creating an array of images 
   \includegraphics[height=5cm]{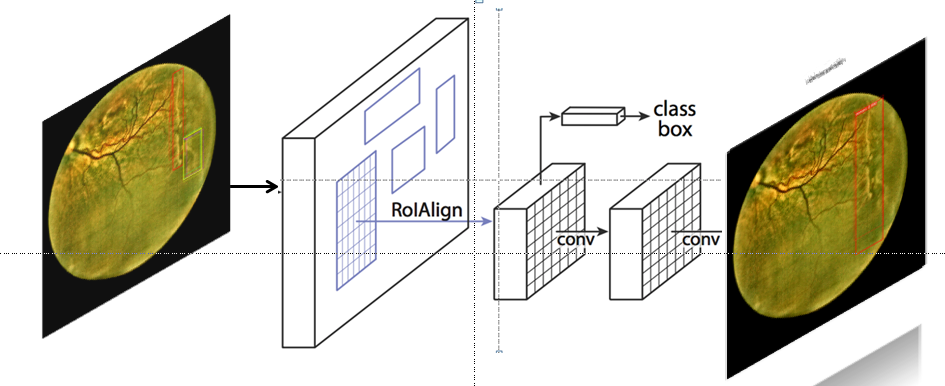}
   \end{tabular}
   \end{center}
   \caption[example] 
%>>>> use \label inside caption to get Fig. number with \ref{}
   { \label{fig:example4} 
Mask R-CNN Framework.}
 \end{figure} 
 
\vspace{0.5cm}
\subsection{Mask RCNN framework}
Framework for Mask RCNN model is shown in Figure. 4. Mask R-CNN extends Faster R-CNN[\citen{Ren}] by adding a branch for predicting segmentation masks on each region of interest (ROI), in parallel with the existing  branch  for  classification  and  bounding  box  regression. Therefore Mask R-CNN can be seen more broadly as a flexible framework for instance-level recognition[\citen{KaimingGD}].

The working principle of Mask R-CNN is quite simple. The model can be roughly divided into 2 parts a region proposal network (RPN) and binary mask classifier. The innovating aspect of Mask R-CNN is that it decouples class prediction and mask generation. The original faster R-CNN has two outputs for class label and a bounding-box offset. Mask R-CNN adds the third branch that outputs the object mask extracting a finer spatial layout of an object. The mask branch is a small fully-connected network applied to each ROI, predicting a segmentation mask in a pixel-to-pixel manner. Because pixel-level segmentation requires much more fine-grained alignment than bounding boxes, mask R-CNN improves the ROI pooling layer (named "ROI Align layer") so that ROI can be better and more precisely mapped to the regions of the original image. The ROI Align network works on principles of object detection, but it outputs multiple possible bounding boxes rather than a single definite one. These boxes are refined using another regression model (Bounding Box Regressor). Afterwards mask prediction is applied separately to each ROI.

In this study, we have used an end-to-end pre-trained Mask R-CNN model with a Resnet-101-FPN backbone. This model has been pre-trained on Imagenet dataset.  It predicts the masks of detected regions and classifies them into one of the classes given at the time of training. We choose an existing open-source implementation [\citen{matterport}] using Tensorflow deep learning framework. We manage to reuse without modification the part of [\citen{matterport}] corresponding to the network, however we had to change the pipeline to preprocess images.

\subsection{Comparison of Mask RCNN with pre-processed image}
To improve generalizablity of the model pre-processing of data is necessary. Also with poor quality neonatal image, enhancement of images plays the part of pre-processing step. If the network is trained without pre-processing of the retinal images, then in some cases ridge will not be localized. Figure. 5 shows the comparison of raw and pre-processed image. The performance of the pre-processing can be evaluated visually to bring out the features in some cases where demarcation line is not at all visible in the input images.

Thus image  pre-processing  is  one  of  the  preliminary step  which  is highly required to ensure sufficient accuracy of the subsequent steps. Training with raw images can not detect ridge in some of the cases where as same image with enhancement is able to detect the ridge correctly.
\begin{figure} [ht]
   \begin{center}
   \begin{tabular}{c} %% tabular useful for creating an array of images 
   \includegraphics[height=7cm]{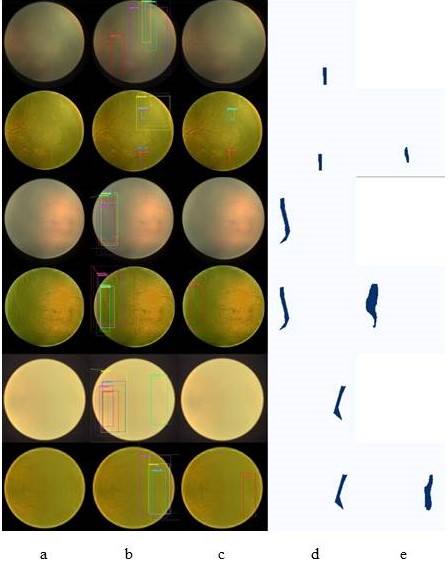}
   \end{tabular}
   \end{center}
   \caption[example] 
%>>>> use \label inside caption to get Fig. number with \ref{}
   { \label{fig:example5} 
Random samples of unsuccessful prediction of masks with raw images. The first, third and fifth row a) contains input neonatal raw images with b) Region of interest (ROI) after refinement, c) detection after non maximum suppression (NMS) d) ground truth mask of image e) predicted mask with this model.
Second, fourth and sixth row with enhanced images after preprocessing.}
 \end{figure}

\section{EXPERIMENT}

To evaluate the model performance, for each image, detections made by the model were compared to the manually determined labels as the ground truth for that image. We randomly choose some qualitative examples of the predicted masks, shown in Figure. 6.

\begin{figure} [ht]
   \begin{center}
   \begin{tabular}{c} %% tabular useful for creating an array of images 
   \includegraphics[height=16cm]{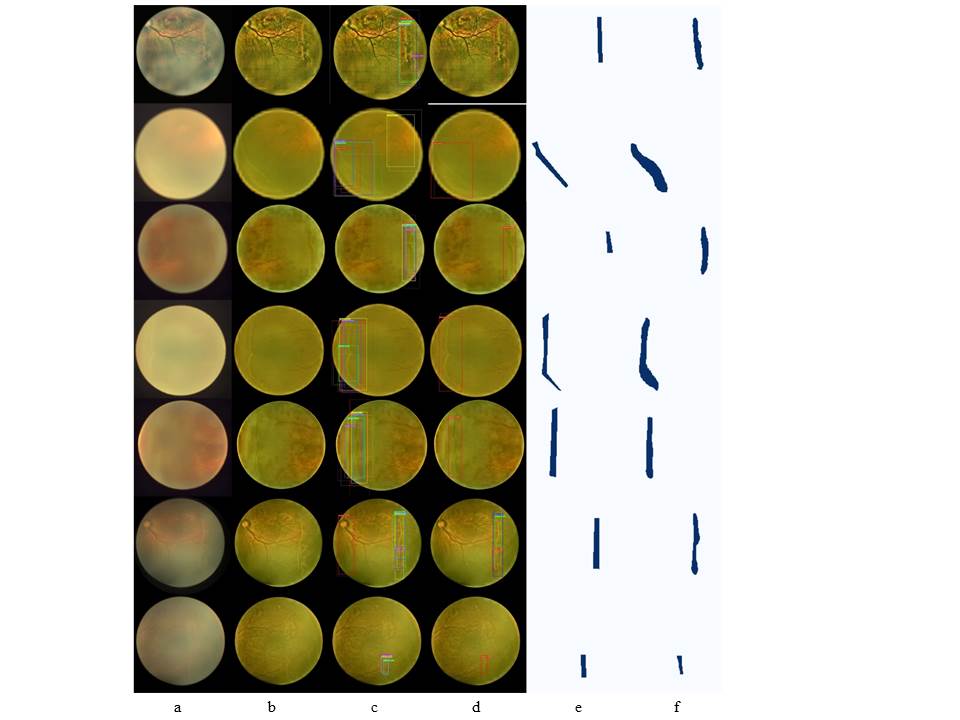}
   \end{tabular}
   \end{center}
   \caption[example] 
%>>>> use \label inside caption to get Fig. number with \ref{}
   { \label{fig:example6} 
Random samples of successfully predicted masks. The first column a) contains input neonatal images. The
Other columns b) enhanced image after preprocessing, c) Region of interest (ROI) after refinement, d) detection after non maximum suppression (NMS) e) ground truth mask of image f) predicted mask with this model. (The figures are best viewed in color.)
}
\end{figure}

For all experiments described here, we use a Mask-RCNN model with a feature pyramid network backbone. For these experiments, we tried ResNet-101 feature pyramid network model as a backbone.  We initialize the model using weights obtained from pre-training on the ImageNet dataset [\citen{imagenet}] and proceed to train the model.  In total we train for 100 epochs using stochastic gradient descent with momentum of 0.9, learning rate of 0.001. 

\subsection{Implementation Setup}
The experimental setup is shown in Table 1. The whole setup was implemented in Linux environment using NVIDIA GTX 1080 8 GB GPU, on a system with 16GB RAM and having Intel core i5 7th generation 3.20 GHz processor. The packages like python 3.6, CUDA 8.0, cuDNN, tensorflow etc., are required for execution of Mask R-CNN code.

\begin{table}[ht]
\caption{Experimental Environments setup.} 
\label{tab:Multimedia-Specifications}
\begin{center}       
\begin{tabular}{|l|l|l|}
\hline
\rule[-1ex]{0pt}{3.5ex}  CPU & Intel (R) Core (TM) i7-8700 @ 3.20GHz  \\
\hline
\rule[-1ex]{0pt}{3.5ex}  GPU & GeForce GTX1080 Ti   \\
\hline
\rule[-1ex]{0pt}{3.5ex}  Main Memory &  16 GB  \\
\hline
\rule[-1ex]{0pt}{3.5ex}  Operating System & Ubuntu 18.04  \\
\hline
\rule[-1ex]{0pt}{3.5ex}  Packages & pyhon 3.6, CUDA 8.0, cudnn, Keras,Tensorflow  \\
\hline 
\end{tabular}
\end{center}
\end{table}

As can be seen from these figures, the learned RCNN model is able to return accurate masks for the ridge.

\subsection{RESULTS}

The performance of the model was measured in terms of precision and recall over the ridge detection in retinal images. Every correct detection is considered a true positive, any wrong detection as a false positive, and any detection missed by the model as a false negative. Our system had F1 score of 0.93, specificity of 0.75, with a precision of 0.97 and a recall of 0.90. The model was able to detect the ridge when the camera light ring effect in image is low.

To compare existing method, the pixel-based evaluation on 30 retcam images [\citen{JoshiM}] average sensitivity, specificity, positive predictive value and negative predictive value achieved were 60.38\%, 99.66\%, 52.77\% and 99.75\% respectively. Whereas using Mask RCNN on KIDROP 45 images average sensitivity, specificity, positive predictive value and negative predictive value achieved were 90\%, 75\%, 97\% and 42\% respectively. 

\section{DISCUSSION}

Accurate detection of ridge/demarcation line is essential for guiding the clinicians for ROP treatment. We showed an effective method to detect the ridges in neonatal images. This method can be used to collect reliable and accurate data required for studies on effect of ridge formation, duration required to treat ROP and period of medical procedures on patients, ROP stage2 treatment , and consequently their outcome, e.g., whether immediate treatment is required or not. Future work includes gathering more data in the neonatal patients to improve the model performance. Another direction for expansion of the model is to use this for accurate segmentation of ridge as well as use of this model for all ROP stage detection.

There are several medical image analysis tasks for which Mask-RCNN based model could be used for medical image segmentation. Examples of this are automatic nucleus segmentation [\citen{Jeremiah}], segmentation of the left ventricle of the heart or liver and tumor segmentation as described in [\citen{patrick}]. Future work will explore the efficacy and performance of Mask-RCNN-based models for other medical image analysis tasks.

\section{CONCLUSION}
In this paper, we presented the benefits of selecting deep convolutional descriptor in object recognition, especially ridge in retinal images. Ridge detection with mask R-CNN achieved an accuracy of 0.88 which is comparable with state-of-the-art.

% References
\bibliography{SPIE} % bibliography data in report.bib
\bibliographystyle{ieeetr}
 
\end{document}